\documentclass[preprint]{aastex}

\slugcomment{Accepted by PASP on May 2, 2006}

\shorttitle{The Anisoplanatic Point Spread Function}
\shortauthors{Britton}

\begin{document}


\title{The Anisoplanatic Point Spread Function in Adaptive Optics}


\author{M. C. Britton}
\affil{California Institute of Technology, Pasadena, CA 91125}
\email{mbritton@astro.caltech.edu}

\begin{abstract}
The effects of anisoplanatism on the adaptive optics point spread
function are investigated.  A model is derived that combines
observations of the guide star with an analytic formulation of
anisoplanatism to generate predictions for the adaptive optics point
spread function at arbitrary locations within the field of view.  The
analytic formulation captures the dependencies of anisoplanatism on
aperture diameter, observing wavelength, angular offset, zenith angle
and turbulence profile.  The predictions of this model are compared to
narrowband 2.12 $\mu$m and 1.65 $\mu$m images of a 21 arcsec binary
($m_{v}$=7.3, 7.6) acquired with the Palomar Adaptive Optics System on
the Hale 5 meter telescope.  Contemporaneous measurements of the
turbulence profile made with a DIMM/MASS unit are used together with
images of the primary to predict the point spread function of the
binary companion.  Predicted companion Strehl ratios are shown to
match measurements to within a few percent, whereas predictions based
on the isoplanatic angle approximation are highly discrepant.  The
predicted companion point spread functions are shown to agree with
observations to 10\%.  These predictions are used to measure the
differential photometry between binary members to an accuracy of 1
part in $10^{3}$, and the differential astrometry to an accuracy of 1
mas.  Errors in the differential astrometry are shown to be dominated
by differential atmospheric tilt jitter.  These results are compared
to other techniques that have been employed for photometry,
astrometry, and high contrast imaging.
\end{abstract}





\section{Introduction}

An adaptive optics system senses phase aberrations arising from
atmospheric turbulence using observations of a guide star, and
compensates these aberrations by applying a correction to an adaptive
mirror.  This compensation is valid in the direction of the guide
star, but degrades with angular offset from the guide star due to
anisoplanatism.  This effect arises due to the shearing between the
columns of turbulent atmosphere traversed by light from the guide star
and light from a target at a finite angular offset.

The phase aberrations that arise from anisoplanatism depend on a
number of parameters.  The errors grows with angular offset from the
reference source, so that image quality and Strehl ratio degrade with
increasing angular offset.  The vertical distribution of turbulence
has a strong effect on the degree of error, with higher altitude
turbulence generating larger errors due to the larger geometrical
shear.  This dependence varies in time as the distribution of
atmospheric turbulence over the telescope evolves.  The anisoplanatic
error grows with zenith angle, since one sees more turbulence along
line of sight to the object.  The error is also a strong function of
aperture diameter and observing wavelength.

This large number of dependencies yields a rich phenomenology for the
adaptive optics point spread function (PSF).  The parameter space is
so large that it is not practical to integrate long enough to attain
the stochastically averaged PSF.  The variability in the adaptive
optics PSF is a serious impediment to the quantitative interpretation
of observations, and can limit the precision in astronomical
applications involving photometry, astrometry, crowded field imaging,
and high dynamic range imaging of extended objects.  Integral field
unit spectroscopy of extended objects is another application in which
the dynamic range of observations may be seriously compromised by PSF
variability.

Multiconjugate \citep{1988vltt.conf..693B} and multiobject
\citep{2004SPIE.5382..727H, 2005SPIE.5903...20E} adaptive optics
architectures have been proposed that aim to directly overcome the
effects of anisoplanatism.  These architectures use multiple guide
stars distributed over a finite field of view and employ tomographic
algorithms to estimate the three dimensional volume of atmospheric
turbulence.  The algorithms rely on a knowledge of the angular offsets
among the guide stars, and use the statistical correlations induced by
these angular offsets in order to effect the turbulence estimation.
In this sense, these algorithms employ anisoplanatism in order to
overcome the effects of anisoplanatism.  Anisoplanatism in a single
conjugate adaptive optics system forms the limiting case of this more
complex problem.  On-sky tests with existing adaptive optics systems
can serve an important role in validating our understanding of
anisoplanatism and increasing our confidence in the success of
tomography.

Significant efforts have been directed towards overcoming the effects
of anisoplanatism in existing adaptive optics systems.  One line of
investigation has aimed to extract estimates of the PSF from observed
data.  For a small target field, the PSF may be assumed to be field
independent.  Observations of such fields that contain multiple point
sources may be deconvolved using a reference PSF selected from the
data or by solving for the optimal PSF during deconvolution.  These
approaches have been applied to crowded stellar fields
\citep{diolaiti00b, christou04} and planetary objects
\citep{2004Icar..169..250D}.  In an effort to calibrate wider fields,
\cite{steinbring02} measured the effects of anisoplanatism on the PSF
using observations of crowded fields, and then applied these
measurements to other fields of interest.  This type of technique is
unable to capture any temporal evolution of the turbulence profile
that occurs between the observations.  Another approach has employed a
parameterized model of the effects of anisoplanatism on the adaptive
optics PSF, extracting these parameters during the deconvolution
procedure \citep{flicker05}.

The above techniques did not attempt to use any independent
information about the turbulence profile in estimating the adaptive
optics PSF.  Other researchers have incorporated such measurements
into the PSF estimates.  \cite{1998A&AS..133..427V} computed the
structure function due to residual phase aberrations from
anisoplanatism, and used this to evaluate the Strehl ratio as a
function of angular offset from the guide star for different
turbulence profiles and aperture diameters.
\cite{2000A&AS..142..149F} derived an expression for the adaptive
optics optical transfer function that captured the depenedencies of
anisoplanatism on the above parameters.  These authors carried out 850
nm observations of two binaries with the ONERA adaptive optics system
and used a turbulence profile measured from a balloon flight together
with the analytical expression for the OTF to predict the binary
companion PSF.  \cite{weiss2002} performed K band adaptive optics
observations of a binary using the ALFA adaptive optics system while
at the same time measuring the turbulence profile using a scidar
instrument.  These authors compared the Strehl ratio degradation due
to anisoplanatism measured from the binary image data with that
expected from the measured turbulence profiles.  These experiments
have shown a promising level of agreement between predictions and
measurements.

Recent work in the area of turbulence monitoring has yielded an
automated set of equipment capable of delivering real time estimates
of the turbulence profile on minute timescales
\citep{2005PASP..117..395T}.  This equipment is based on the
combination of a Multi-Aperture Scintillation Sensor (MASS)
\citep{2003SPIE.4839..837K} and a Differential Image Motion Monitor
(DIMM) \citep{1995PASP..107..265V}.  This turbulence monitoring
equipment is now being used at a number of different sites, and has
been employed by the Thirty Meter Telescope project for its site
testing program \citep{2004SPIE.5489..154S}.  As part of this program,
a set of this equipment has been installed at Palomar Observatory.

This paper describes an experiment in which measurements of the
turbulence profile from the DIMM/MASS equipment and short exposure
images of a 21 arcsec binary from the Palomar Adaptive Optics
System on the Hale 5 meter telescope were acquired contemporaneously
over the course of several hours.  Section \ref{theory} presents the
analysis of anisoplanatism used for this experiment.  Section
\ref{observations} describes the observations, while Section
\ref{data_analysis} and \ref{results} describe the analysis and
results of the experiment.

\section{A Model for the Field Dependent AO PSF}
\label{theory}

\subsection{Model Formulation}

Consider an adaptive optics system that measures the wavefront phase
aberrations $\phi_{a}(\vec{r})$ in the direction of a guide star and
compensates these aberrations using an adaptive mirror.  Here
$\vec{r}$ is a vector in the pupil plane of the telescope.  An
adaptive optics system is not capable of effecting a perfect
correction, and residual phase aberrations
$\tilde{\phi}_{a}\left(\vec{r}\right)$ will be present in the guide
star wavefront after adaptive compensation.  The wavefront phase
aberrations $\phi_{b}(\vec{r})$ in another direction on the sky differ
from those of the guide star due to anisoplanatism.  In this
direction, the residual aberrations that remain after compensation by
the adaptive optics system are
\begin{eqnarray}
\label{resid_phase}
\Delta\phi\left(\vec{r}\right) & = & \phi_{b}\left(\vec{r}\right) - \phi_{a}\left(\vec{r}\right) + \tilde{\phi}_{a}\left(\vec{r}\right)
\end{eqnarray}
This simple model does not account for time delay in the application
of the adaptive correction.  During this delay the wind carries
turbulence past the telescope aperture, causing a servo error in the
guide star wavefront.  This turbulence evolution also induces
correlations between the adaptive correction and the wavefront
aberrations in other directions on the sky \citep{tyler83}.  Under
certain observing conditions, this effect can lead to a situation
where the Strehl ratio can be higher downwind of the guide star
\citep{1999aoa..book..351L}.  In this analysis, this wind induced
anisoplanatic effect is neglected.

The structure function for a random process $\psi(\vec{r})$ is
\begin{eqnarray}
\label{strfn_def}
D_{\psi}(\vec{r}_{1},\vec{r}_{2}) & = & 
\left\langle \left\{\psi\left(\vec{r}_{1}\right) - 
\psi\left(\vec{r}_{2}\right)\right\}^{2}\right\rangle
\end{eqnarray}
Using Equation \ref{resid_phase}, one can write the structure function
for the residual phase $\Delta\phi\left(\vec{r}\right)$ as
\begin{eqnarray}
\label{strfn}
D_{\Delta\phi}(\vec{r}_{1},\vec{r}_{2}) & = & 
D_{\rm apl}(\vec{r}_{1},\vec{r}_{2}) + D_{\tilde{\phi}_{a}}(\vec{r}_{1},\vec{r}_{2}) + \\
& & 2 
\left\langle 
\left\{
\phi_{\rm apl}(\vec{r}_{1})\tilde{\phi}_{a}(\vec{r}_{1}) +
\phi_{\rm apl}(\vec{r}_{2})\tilde{\phi}_{a}(\vec{r}_{2}) -
\phi_{\rm apl}(\vec{r}_{1})\tilde{\phi}_{a}(\vec{r}_{2}) -
\phi_{\rm apl}(\vec{r}_{2})\tilde{\phi}_{a}(\vec{r}_{1}) 
\right\}
\right\rangle \nonumber
\end{eqnarray}
where $\phi_{\rm apl}(\vec{r}) = \phi_{b}\left(\vec{r}\right) -
\phi_{a}\left(\vec{r}\right)$ is the component of the residual phase
arising from anisoplanatism.  The four cross terms in Equation
\ref{strfn} represent statistical correlations between the residual
phase errors from the adaptive optics correction and those from
anisoplanatism.  A number of error terms that arise in adaptive optics
systems have no correlation with anisoplanatic errors.  Examples
include measurement errors, aberrations in the optical system, and
non-common path errors.  In contrast, residual fitting error and servo
error are weakly correlated with anisoplanatic errors.  While the
relative importance of these cross terms will depend on the error
budget of the AO system, in many circumstances these terms will be
small relative to the structure functions in Equation \ref{strfn}.
Here, these cross terms are neglected.

With this approximation, the optical transfer function may be written
as \citep{1985stop.book.....G}
\begin{eqnarray}
\label{otf_eqn}
OTF(\vec{r}) & = & \int d\vec{s} 
\exp{\left\{-{1 \over 2}\left[
D_{\rm apl}(\vec{s},\vec{r} + \vec{s}) + 
D_{\rm \tilde{\phi}_{a}}(\vec{s},\vec{r} + \vec{s})\right]\right\}}W(\vec{s})W(\vec{r}+\vec{s}) 
\end{eqnarray}
where $W(\vec{r})$ is the pupil function.  An issue of particular
importance in evaluating the OTF is whether the structure functions of
the two random processes are stationary over the pupil plane.  For
such a process, $D_{\psi}(\vec{r}_{1},\vec{r}_{2}) =
D_{\psi}(\vec{r}_{2} - \vec{r}_{1})$.  As will be shown in the next
section, $D_{\rm apl}(\vec{r}_{1},\vec{r}_{2})$ is in fact stationary.
This permits the OTF to be written as
\begin{eqnarray}
\label{factored_otf}
OTF(\vec{r}) & = & 
\exp{\left\{-{1 \over 2}
D_{\rm apl}(\vec{r})\right\}} 
\int d\vec{s} 
\exp{\left\{-{1 \over 2}
D_{\rm \tilde{\phi}_{a}}(\vec{s},\vec{r} + \vec{s})\right\}}W(\vec{s})W(\vec{r}+\vec{s})  
\end{eqnarray}
Note that the OTF has factored into a term that describes the
anisoplanatic errors and a term that describes the residual errors in
the direction of the guide star.

A relationship similar to Equation \ref{factored_otf} was derived by
\cite{2000A&AS..142..149F}, and represents a very important result.
One may formulate the OTF in any direction on the sky from the product
of the guide star OTF and an anisoplanatic transfer function formed
from $D_{\rm apl}(\vec{r})$.  This factorization is of considerable
use in the formulation of a practical scheme for evaluating the OTF,
as discussed below.

\subsection{The Anisoplanatic Structure Function}
\label{aniso_math}

Prediction of the OTF in Equation \ref{factored_otf} requires an
evaluation of the anisoplanatic structure function $D_{\rm
apl}(\vec{r})$.  This function may be computed using a semianalytic
expression for the piston removed phase covariance on a circular
aperture in the presence of Komolgorov turbulence \citep{tyler94}.
The covariance between the piston removed wavefront phase from two
different sources at two different points $\vec{r}_{1}$ and
$\vec{r}_{2}$ in the telescope pupil plane is given by
\begin{eqnarray}
\label{covariance}
\left\langle\phi_{a}\left(\vec{r}_{1}\right)\phi_{b}\left(\vec{r}_{2}\right)\right\rangle & =  &
\Xi k^{2} D^{5/3}\int_{0}^{\infty}dz \; C_{n}^{2}\left(z\right) 
\left[
G_{1}\left(\left|{2 \over D}\vec{r}_{1} + \vec{\Omega}_{ab}\left(z\right)\right|\right) +
\right.\\
& & 
\left.
G_{1}\left(\left|{2 \over D}\vec{r}_{2} - \vec{\Omega}_{ab}\left(z\right)\right|\right) - 
\left|{2 \over D}\left(\vec{r}_{1} - \vec{r}_{2}\right) + \vec{\Omega}_{ab}\left(z\right) \right|^{5/3} - 
G_{2}\left(\left|\vec{\Omega}_{ab}\left(z\right) \right|\right)\right] \nonumber
\end{eqnarray}
Here $D$ is the aperture diameter, $k$ is the wavenumber, and
$C_{n}^{2}(z)$ is the turbulence profile along the line of sight to
the star.  The numerical constant $\Xi$ is given by
\begin{equation}
\Xi = {1 \over 5}\left({1 \over 2}\right)^{7/3}\left[\Gamma\left(1 \over 6\right)\right]^{2}\left[\Gamma\left(1 \over 3\right)\right]^{-1} = .458986
\end{equation}
where $\Gamma\left(z\right)$ is the Gamma function.
The quantity $\vec{\Omega}_{ab}\left(z\right)$ is
\begin{equation}
\vec{\Omega}_{ab}\left(z\right) = \left({2 z \over D}\right)\vec{\theta}_{ab}
\end{equation}
where $\vec{\theta}_{ab}$ is the angular offset between the two
sources.  This parameter may be interpreted as the altitude dependent
shear between the two beams.  Finally, the two functions $G_{1}$ and
$G_{2}$ in Equation \ref{covariance} are defined in terms of the Gauss
hypergeometric function
\begin{equation}
\vphantom{F}_{2}F_{1}\left(a, b; c; x\right) = 
\sum_{n=0}^{\infty} {\Gamma\left(a+n\right)\Gamma\left(b+n\right)\Gamma\left(c\right) \over
\Gamma\left(a\right)\Gamma\left(b\right)\Gamma\left(c+n\right)}\;{x^{n} \over n!}
\end{equation}
as
\begin{equation}
G_{1}\left(x\right) = 
\left\{
\begin{array}{ll}
{6 \over 11}\vphantom{F}_{2}F_{1}\left(-{11 \over 6}, -{5 \over 6}; 1; x^{2}\right) & x \le 1 \\
x^{5/3}\vphantom{F}_{2}F_{1}\left(-{5 \over 6}, -{5 \over 6}; 2; x^{-2}\right) & x \ge 1 
\end{array}
\right.
\end{equation}
\begin{equation}
\label{g2_equation}
G_{2}\left(x\right) = 
{8 \over \pi} \int_{0}^{1} dy \left[\cos^{-1}y  - y \sqrt{1-y^{2}}\right]
\left\{
\begin{array}{ll}
\left(2y\right)^{8/3}\vphantom{F}_{2}F_{1}\left(-{5 \over 6}, -{5 \over 6}; 1; \left(x\over 2y\right)^{2}\right) & x \le y \\
2 y x^{5/3}\vphantom{F}_{2}F_{1}\left(-{5 \over 6}, -{5 \over 6}; 2; \left(2y \over x\right)^{2}\right) & x \ge y 
\end{array}
\right.
\end{equation}
Note that the covariance expression contains dependencies on observing
wavelength, aperture diameter, turbulence profile, and angular offset
between the two stars.  The range variable $z$ implicitly incorporates
the dependence of the covariance on zenith angle, which enters through
$C_{n}^{2}(z)$, $\vec{\Omega}_{ab}(z)$, and the range integral itself.

The expression for the covariance in Equation \ref{covariance} is of
great utility, and a number of familiar results in adaptive optics may
be derived from this expression.  Three limiting cases are shown in
Appendix A: the phase structure function in the presence of
uncompensated turbulence, the aperture averaged phase variance in the
presence of uncompensated turbulence, and the aperture averaged
residual phase variance due to anisoplanatism.  These cases motivate
the definition of the Fried parameter $r_{0}$ and the isoplanatic
angle $\theta_{0}$, which are restated in this Appendix.

%%%%%%%%%%%%%%%%
%%% Figure 1 %%%
%%%%%%%%%%%%%%%%
\begin{figure}[t!]
\begin{center}
\includegraphics[width=\textwidth]{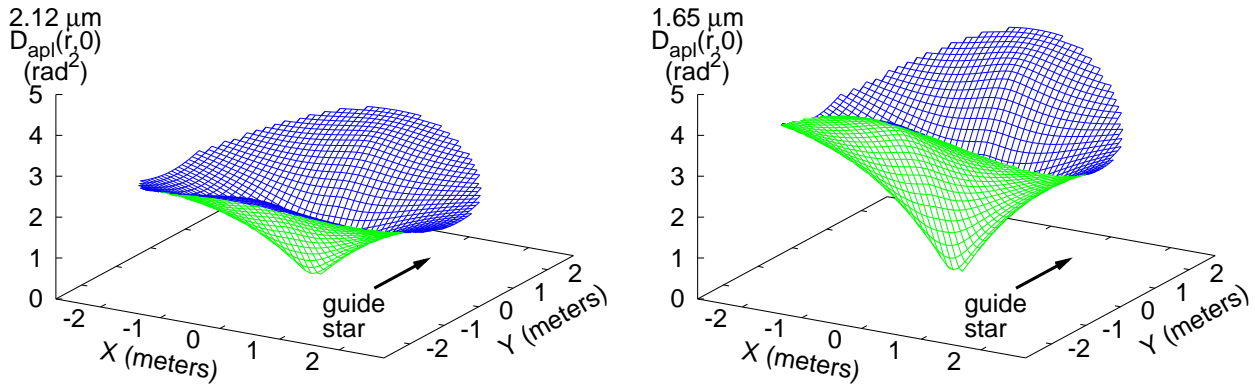}
\end{center}
\figcaption{Anisoplanatic structure functions for a star at a 20
arcsec offset from the guide star, computed using Equation
\ref{aniso_strfn}.  The left and right panels show $D_{\rm
apl}(\vec{r},0)$ at observing wavelengths of 2.12 $\mu$m and 1.65
$\mu$m as a function of location $\vec{r}$ within the pupil plane of a
5 meter telescope.  The angular offset from the guide star was
oriented along the Y axis, and the broken X-Y symmetry gives rise to
the anisotropy apparent in these functions.  This anisotropy is
responsible for the radial elongation of the off axis PSF that is a
characteristic feature of anisoplanatism.  Despite this anisotropy the
structure function is stationary over the pupil plane, so that $D_{\rm
apl}(\vec{r}_{1},\vec{r}_{2}) = D_{\rm
apl}(\vec{r}_{2}-\vec{r}_{1})$. Typical atmospheric turbulence
conditions at Palomar were assumed in these calculations.
\label{strfn_fig}
}
\end{figure}

Expanding the expression for the anisoplanatic structure function
using Equation \ref{strfn_def} permits expression of $D_{\rm
apl}(\vec{r}_{1}, \vec{r}_{2})$ as a sum over ten phase covariance
terms involving $\phi_{a}(\vec{r})$ and $\phi_{b}(\vec{r})$.  Direct
evaluation of the resulting expression using Equation \ref{covariance}
yields 
\begin{eqnarray}
\label{aniso_strfn}
D_{\rm apl}\left(\vec{r}_{1},\vec{r}_{2}\right) & = &
2\Xi k^{2} D^{5/3} \int_{0}^{\infty} dz \; C_{n}^{2}(z) 
\left\{
2 \left\vert\vec{\Omega}_{ab}\right\vert^{5/3} +
2 \left\vert {2 \over D} \left(\vec{r}_{1} - \vec{r}_{2}\right)\right\vert^{5/3} - 
\right.
\\
& & 
\left.
\left\vert {2 \over D} \left(\vec{r}_{1} - \vec{r}_{2}\right) + \vec{\Omega}_{ab}\right\vert^{5/3} - 
\left\vert {2 \over D} \left(\vec{r}_{1} - \vec{r}_{2}\right) - \vec{\Omega}_{ab}\right\vert^{5/3}
\right\}
\nonumber
\end{eqnarray}
This result indicates that the anisoplanatic structure function
$D_{\rm apl}\left(\vec{r}_{1},\vec{r}_{2}\right)$ is in fact only a
function of $\vec{r}_{1} - \vec{r}_{2}$, and is therefore stationary
over the pupil plane.  This property permits the factorization of the
anisoplanatic and guide star OTFs shown in Equation
\ref{factored_otf}.

As in the case of the covariance, $D_{\rm apl}\left(\vec{r}\right)$
depends on observing wavelength, aperture diameter, turbulence
profile, zenith angle and angular offset between the guide star and
the direction of interest.  Examples of the anisoplanatic structure
function are shown in Figure \ref{strfn_fig} for a particular set of
these parameters.

\subsection{The Anisoplanatic OTF and PSF}
\label{aniso_numerical}

%%%%%%%%%%%%%%%%
%%% Figure 2 %%%
%%%%%%%%%%%%%%%%
\begin{figure}[t!]
\begin{center}
\includegraphics[width=\textwidth]{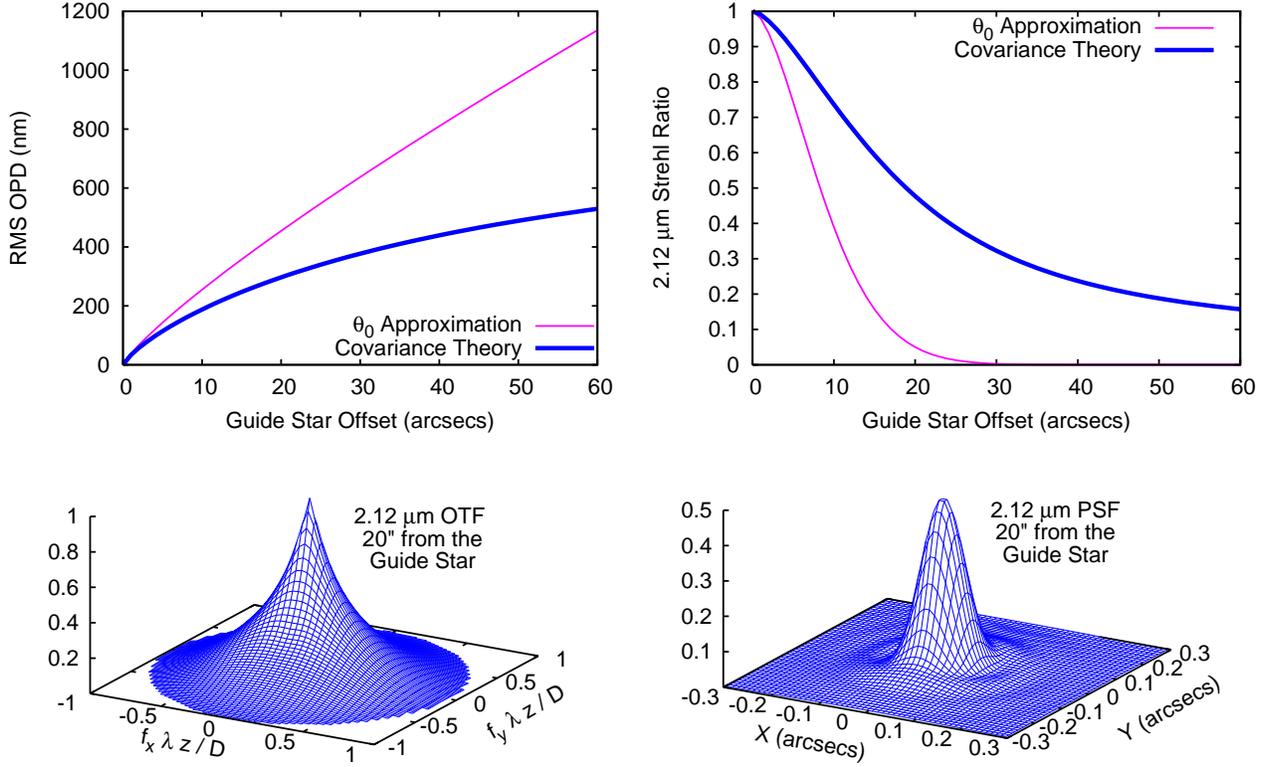}
\end{center}
\figcaption{Illustrative effects of ansioplanatism on a 5 meter
telescope, as described in the text.  The upper left panel shows the
aperture averaged residual RMS OPD due to anisoplanatism vs angular
offset.  The upper right panel shows the 2.12 $\mu$m Strehl ratio vs
angular offset.  The lower panels show the 2.12 $\mu$m anisoplanatic
OTF and PSF at an angular offset of 20 arcsecs from the guide star.
The PSF has been normalized by the peak value of the ideal,
diffraction limited PSF.  Typical atmospheric turbulence conditions
for Palomar were assumed for these calculations.
\label{aniso_fig}
}
\end{figure}

Equation \ref{factored_otf} represents a model for the OTF at any
point in the field of view.  This quantity is a strong function of
many parameters, which enter through both the anisoplanatic structure
function $D_{\rm apl}\left(\vec{r}\right)$ and the
guide star OTF.  These dependencies give rise to the rich morphology
in the adaptive optics PSF that presents such a challenge in the
analysis of adaptive optics observations.

The classic dependence displayed by the adaptive optics PSF is its
degradation with increasing angular offset from the guide star.
Figure \ref{aniso_fig} contains a series of plots that illustrate this
behavior for a 5 meter telescope under typical atmospheric turbulence
conditions at Palomar.  The first plot shows the aperture averaged
residual RMS optical path difference (OPD) vs angular offset from the
guide star.  These results were computed by evaluating
$\left\langle\left[\phi_{\rm apl}(\vec{r})\right]^{2}\right\rangle$
using Equation \ref{covariance}, and averaging this quantity over the
pupil plane.  The OPD is found by computing the square root of this
result and dividing by $k$.  Also shown in this plot is the OPD
computed from the $\theta_{0}$ approximation described in Appendix A.
Even at small angular offsets, significant discrepancies exist between
the exact result and the $\theta_{0}$ approximation.

Consider an idealized adaptive optics system that acts to eliminate
all wavefront error in the direction of the guide star, so that
$\tilde{\phi}(\vec{r})=0$.  In this case the OTF in Equation
\ref{factored_otf} depends only on the anisoplanatic structure
function $D_{\rm apl}\left(\vec{r}\right)$, which may be evaluated
using Equation \ref{aniso_strfn}.  This allows computation of an OTF
that incorporates only the effects of anisoplanatism, from which a PSF
may be evaluated by Fourier transformation.  Figure \ref{aniso_fig}
shows the 2.12 $\mu$m anisoplanatic OTF and PSF at an angular offset
of 20 arcsecs from the guide star, computed using this method.  The
PSF displays an increase in elongation along the direction to the
guide star that is a characteristic of anisoplanatism.  At this
angular offset, the Strehl ratio has dropped to 47\% due solely to the
effects of anisoplanatism.

In the same way, the anisoplanatic Strehl ratio may be computed as a
function of angular offset from the guide star.  These results are
also plotted in Figure \ref{aniso_fig}.  For comparison, the
anisoplanatic Strehl ratios computed from the Marechal approximation
are also shown.  In this approximation, the aperture averaged phase
variance computed from the $\theta_{0}$ approximation was
exponentiated to form an estimate of the anisoplanatic Strehl ratio.
The combination of these two approximations significantly
underestimates the anisoplanatic Strehl ratio computed directly from
the anisoplanatic PSF.

This approach to evaluating the anisoplanatic OTF and PSF is
illustrative, but the application of Equation \ref{factored_otf} to
observational data requires a treatment of the guide star OTF.  While
significant efforts have been made to estimate the guide star OTF
using the statistical properties of $\tilde{\phi}_{a}(\vec{r})$
\citep{veran97}, evaluation of Equation 5 is considerably simplified
if an observation of the guide star may be used in the prediction of
the PSF elsewhere in the field of view.  For this procedure to work,
the guide star must itself be a point source.  This is often the case
in adaptive optics observations.  In these circumstances, the guide
star PSF may be extracted from the observational data and Fourier
transformed to form the guide star OTF that appears in Equation
\ref{factored_otf}.  The anisoplanatic structure function $D_{\rm
apl}\left(\vec{r}\right)$ may be evaluated from Equation
\ref{aniso_strfn} for the field point of interest.  The only parameter
in Equation \ref{aniso_strfn} that is not defined by the observation
is the turbulence profile, which may be measured using an independent
set of equipment.  The product of the observed guide star OTF and the
anisoplanatic transfer function provides a prediction of the OTF at
the field point of interest.  Fourier transformation of this OTF
yields a prediction for the PSF.  There are no free parameters in this
prediction.

\section{Observations}
\label{observations}

To carry out a comparison of the model in Equation \ref{factored_otf}
with AO compensated image data, a set of observations was carried out
at Palomar Observatory on the night of August 20, 2005.  In this
experiment, a well separated binary system was chosen and one member
of the binary was used as the guide star for the Palomar Adaptive
Optics system \citep{2000SPIE.4007...31T} on the Hale 5 meter
telescope.  A sequence of AO compensated near infrared images of the
binary system from the PHARO infrared camera
\citep{2001PASP..113..105H} and a sequence of turbulence profiles from
a set of DIMM/MASS equipment were acquired contemporaneously.  Using
the methodology presented in Section \ref{theory}, a prediction for
the PSF of the binary companion was formulated from the observed PSF
of the guide star, the measured turbulence profiles, and the
parameters defined by the observation.  This prediction was then
compared directly to the observed images of the binary companion.
This section describes the details of these observations.

\subsection{Turbulence Profile Measurements}

%%%%%%%%%%%%%%%%
%%% Figure 3 %%%
%%%%%%%%%%%%%%%%
\begin{figure}[t!]
\begin{center}
\includegraphics[width=\textwidth]{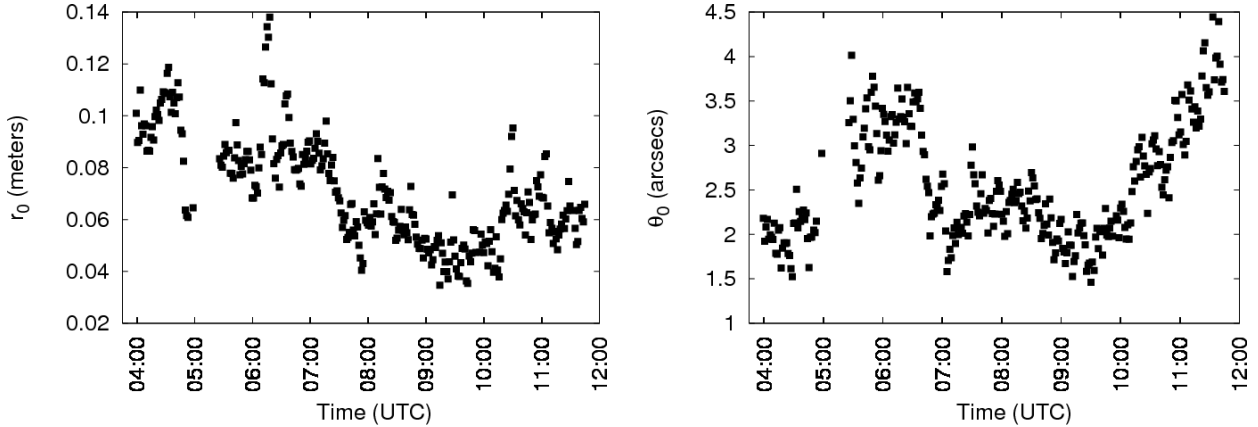}
\end{center}
\figcaption{Turbulence parameters on the night of August 20, 2005
measured using the DIMM/MASS equipment at Palomar.  These parameters
were computed from the seven layer profiles estimated from the
DIMM/MASS measurements, and are quoted at zenith and at a reference
wavelength of .5 $\mu$m.  The mean value of the Fried parameter
$r_{0}$ was 7.1 cm, and that of the isoplanatic angle $\theta_{0}$ was
2.54 arcsecs. Both parameters displayed considerable excursions
from their means over the course of the night.  
\label{turbulence_params_fig}
}
\end{figure}

Turbulence profile measurements were acquired using a DIMM/MASS unit
on loan from the Thirty Meter Telescope Project.  The unit itself has
been described by \cite{2004SPIE.5489..154S}, and consists of a 35 cm
robotic telescope that feeds both a DIMM and MASS instrument.  The
methodology employed in deriving turbulence profiles from the
combination of DIMM and MASS measurements is discussed by
\cite{2003MNRAS.340...52T} and references therein.  The DIMM measures
the differential motion between images acquired through two adjacent
apertures.  This measurement is sensitive to the integrated turbulence
profile.  A MASS unit measures the scintillation index of images
acquired through four apertures of different radii.  These
measurements are sensitive to the distribution of higher altitude
atmospheric turbulence, which gives rise to scintillation.  Together
these measurements permit an estimate of the turbulence profile at
altitudes of 0, .5, 1, 2, 4, 8 and 16 km.  In a comparison of the
turbulence strength measured at each of these layers, simultaneous
MASS and scidar observations on Mauna Kea have displayed agreement to
a factor of several \citep{2005PASP..117..395T}.  One of the
difficulties in the reconstruction of the turbulence profile from MASS
data is that the procedure is somewhat ill-conditioned, in that
simultaneously adjusting the layer altitude and the strength of
turbulence can lead to similar scintillation indices.  However, to
leading order anisoplanatism displays this same type of covariance
({\it c.f.}  Equation A13).  In this sense both anisoplanatism and
MASS are low resolution atmospheric profilometers.

The DIMM/MASS at Palomar Observatory was located 300 meters from the
Hale 5m telescope.  This unit was installed on the roof of a building
approximately 10m above ground level.  Turbulence profiles were
measured about once every 90 seconds throughout the night of August
20th.  For the first hour of the experiment, the DIMM/MASS unit
tracked Beta Drac at $17^{h}30^{m} +52^{\circ}18''$.  A half hour gap
in coverage occurred between 5:00 and 5:30 UT due to a tracking error
experienced by the robotic telescope.  From 5:30 UT onwards, the unit
recorded images from Alpha Cep at $21^{h}19^{m} +62^{\circ}35''$.  For
comparison, the binary imaged using the AO system was at $17^{h}59^{m}
+64^{\circ}08''$.

 Figure \ref{turbulence_params_fig} shows the values of $r_{0}$ and
$\theta_{0}$ computed from these profiles.  Both of these parameters
display deviations from their means of order a factor of two over the
course of the night.  The variability in the turbulence profiles imply
a factor of two variation in the RMS OPD due to anisoplanatism.  This
in turn yielded pronounced variability in the Strehl ratio of the
binary companion that was readily detectable in the AO compensated
image data.

\subsection{Adaptive Optics Observations}

Adaptive optics observations of the binary system HD164983 + HD164984
were acquired over a three hour period between 4:17 and 7:18 UT.  This
binary system has an angular separation of 21'' and is oriented at
$282^{\circ}$ east of north.  The Johnson V magnitudes of HD164983 and
HD164984 are 7.6 and 7.3, respectively, and the latter source was used
as the guide star for the adaptive optics system.  The 25 arcsec
field of view of the PHARO camera was used, permitting both binary
members to be positioned on the detector simultaneously.  Imaging was
performed using two narrowband filters: an H2 filter with a central
wavelength of 2.123 $\mu$m and a bandpass of .007 $\mu$m and an FeII
filter with a central wavelength of 1.648 $\mu$m and a bandpass of .03
$\mu$m.  Exposure times of 2.8 seconds and 1.4 seconds were chosen for
the H2 and FeII observations, respectively.  The resulting stellar
images peaked at less than 30\% of the detector full well depth, which
is well within the linear range of the detector.  Images were acquired
in a 7 point linear dither pattern, in which the binary was shifted up
and down the detector in 4.5'' steps.  Ten exposures were acquired at
each dither position.  After culling the data for bad images, a total
of 703 exposures at 2.123 $\mu$m and 384 exposures at 1.648 $\mu$m
remained.

Calibration of the image data was carried out in the customary way.
Flat field calibration was performed using twilight sky flats.  Sky
subtraction was performed by forming the median of the dithered
exposures and subtracting this median from each exposure.  Finally,
3.8'' subimages centered on each of the two binary members were
extracted from each exposure for use in the analysis below.  The
angular extent of these subimages was chosen to encompass all residual
scattered light.

\section{Data Analysis}
\label{data_analysis}

A prediction of the PSF at the location of the binary companion was
formulated for each exposure using the method described in Section
\ref{theory}.  Each subimage of the guide star was Fourier transformed
to form the guide star OTF.  An anisoplanatic OTF was computed for
each measured turbulence profile using the observing wavelength,
aperture diameter, angular offset to the binary companion, and zenith
angle at the time of the measurement.  For each adaptive optics
exposure, an anisoplanatic OTF was formed for the time of the exposure
by interpolating the anisoplanatic OTFs computed for the nearest
turbulence profile measurements.  The pointwise product of the guide
star OTF and the interpolated anisoplanatic OTF was formed, and the
resulting OTF was Fourier transformed to generate a PSF prediction for
the binary companion sampled at the 25 mas pixel scale of the PHARO
image data.

For each exposure, the Strehl ratios of each observed binary member
and of the predicted companion PSF were calculated.  To compute these
Strehl ratios, an ideal, diffraction limited PSF for the Hale 5m was
simulated by forming the pupil plane wavefront, Fourier transforming
to form the image plane wavefront, and computing the square modulus.
The pupil plane wavefront used in this procedure accounted for the
shadows cast by the secondary mirror and the four struts that support
it.  For each subimage this diffraction limited PSF was normalized to
have the same integrated signal, and the Strehl ratio was computed as
the ratio of the peak value in the image to that of the ideal PSF.  In
these observations the 2.12 $\mu$m images were oversampled at 1.7
times Nyquist, while the 1.65 $\mu$m images were oversampled at 1.3
times Nyquist.  This oversampling mitigates many of the subtle effects
that can arise in computing Strehl ratios from image data
\citep{2004SPIE.5490..504R}.

In addition to the Strehl ratio analysis, a direct comparison was
carried out between the predicted and observed PSFs of the binary
companion.  Because the guide star PSF was used in formulating this
prediction, the predicted PSF represents the image that would be
obtained for a point source with the same brightness as the guide
star.  In fact the binary companion has a different brightness, which
may be varying in time.  Possible sources of such variability include
the presence of cirrus clouds during observations or intrinsic
photometric variability of the star itself.  In addition, subimages
were extracted from the PHARO exposures at a specific angular offset.
Variations in the binary offset may occur from one exposure to the
next, and it is of interest to measure these offsets from the data.
These parameters constitute the differential photometry and astrometry
of the binary, and their measurement requires a fit of the predicted
companion PSF to the observed companion PSF.

For this analysis, a simple four parameter model was used.  The
relationship between an observation of the binary companion $P(x,y)$
and the predicted PSF $R(x, y)$ was taken to be
\begin{equation}
\label{psf_fit}
P(x,y) = b R(x-\Delta x, y-\Delta y) + c
\label{spatial_model}
\end{equation}
Here $b$ represents the differential amplitude between the two binary
members, while $\Delta x$ and $\Delta y$ represent the differential
angular offsets.  The constant $c$ models any differential in the
background level of the guide star and companion data, and guarantees
that the residuals from the fit have zero mean.  Solution of this
equation is readily performed using the predicted and observed
companion OTFs.  In OTF space, the parameters $\Delta x$ and $\Delta
y$ induce a phase slope by the shift theorem
\citep{1986ftia.book.....B}.  A $\chi^{2}$ merit function is readily
formulated in this space, and minimization of this function leads to a
simple iterative solution for all four parameters.

The outcome of this fitting procedure permits measurement of the
differential astrometry and photometry between binary members.  The
overall angular offsets assumed in the subimage extraction were added
to the fitted values of $\Delta x$ and $\Delta y$ to yield the
differential astrometric offset between the binary members.  The
differential photometry was computed as the ratio of the total flux of
the fitted PSF to that of the guide star.  Because the residuals are
zero mean, the total flux of the fitted PSF is identical to that of
the observed image of the binary companion.  At first glance, the use
of the fitted PSF appears to provide no advantage over the observed
PSF.  However, these images were oversampled and their OTFs were
lowpass filtered to suppress spatial frequencies larger than the
cutoff set by the telescope aperture.  This filtering step
substantially reduces the noise in the image.

\section{Results}
\label{results}

%%%%%%%%%%%%%%%%
%%% Figure 4 %%%
%%%%%%%%%%%%%%%%
\begin{figure}[t!]
\begin{center}
\includegraphics[width=\textwidth]{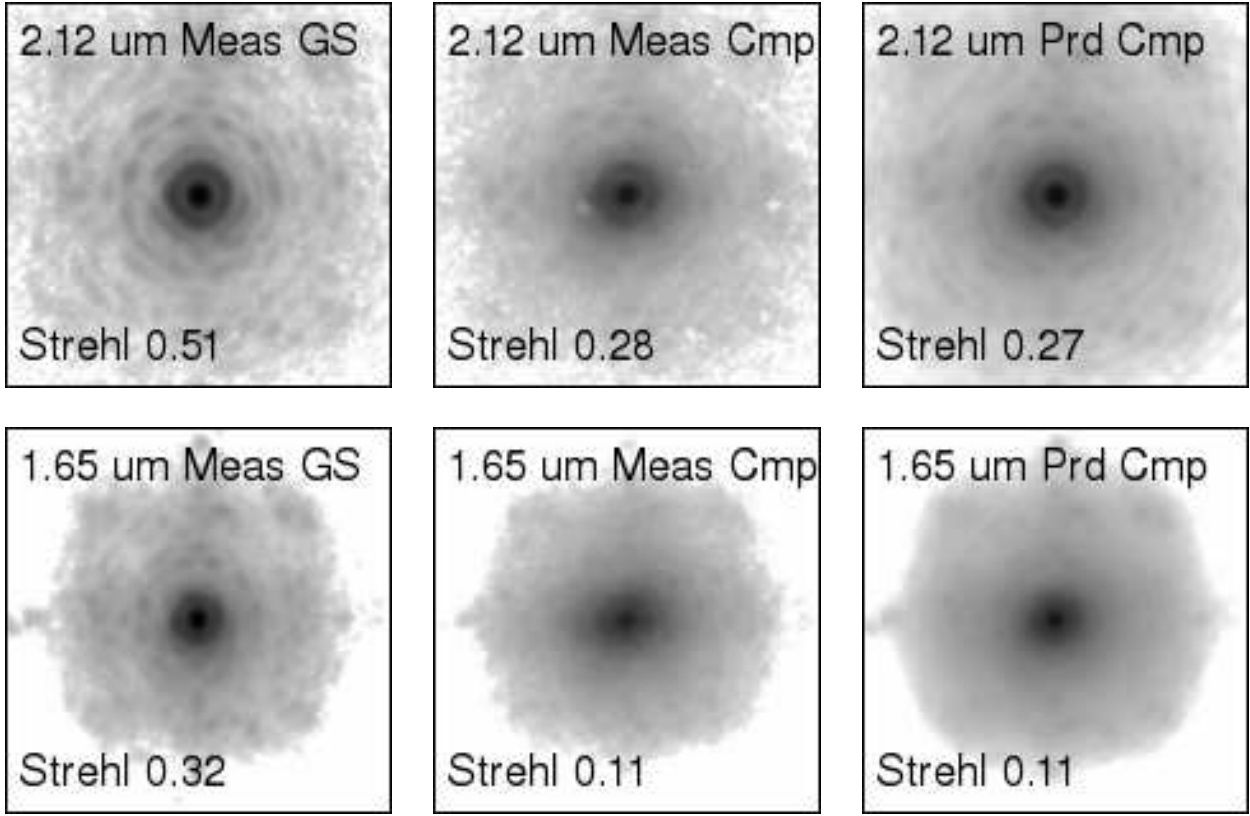}
\end{center}
\figcaption{Observations of the guide star HD164984 and its companion
HD164983 at 2.12 $\mu$m and 1.65 $\mu$m.  These observations were
formed by integrating five sequential exposures, and are displayed on
a log stretch.  The angular offset between these stars is 21.3
arcsecs, and only a 2 arcsec field surrounding each star is
shown.  The first and second columns show observations of the guide
star and companion, respectively.  The predicted PSF of the companion
appears in the third column, and was formulated from the product of
the guide star OTF and the anisoplanatic OTF as described in Section
2.  The Strehl ratios are shown at the bottom of each image.  In these
data, anisoplanatism has degraded the Strehl ratio of the companion by
factors of two to three relative to the guide star.  This degradation
is accurately captured in the predicted companion PSF.
\label{obs_fig}
}
\end{figure}

Examples of the 2.12 $\mu$m and 1.65 $\mu$m observations of the binary
members are shown in Figure \ref{obs_fig}.  These observations were
integrated over five sequential exposures, and serve to illustrate the
quality of the AO compensation delivered by the PALAO system.  In the
particular 14 second observation at 2.12 $\mu$m shown in this figure,
the guide star Strehl ratio was 51\%.  Due to the effects of
anisoplanatism, the measured Strehl ratio of the binary companion was
degraded to 28\%.  The predicted companion PSF computed from the guide
star OTF and the turbulence profile at the time of this observation is
also shown in this figure.  The Strehl ratio calculated for the
predicted companion PSF was 27\%.  The level of agreement between
measured and predicted companion Strehl is a strong indication that
the OTF model in Equation \ref{factored_otf} accurately captures the
effects of anisoplanatism.  This also implies that the turbulence
profiles measured by the DIMM/MASS equipment and used in this model
reflect the true distribution of atmospheric turbulence.  A similar
level of agreement is seen for the 7 second observation at 1.65 $\mu$m
shown in the same figure.

%%%%%%%%%%%%%%%%
%%% Figure 5 %%%
%%%%%%%%%%%%%%%%
\begin{figure}[t!]
\begin{center}
\includegraphics[width=\textwidth]{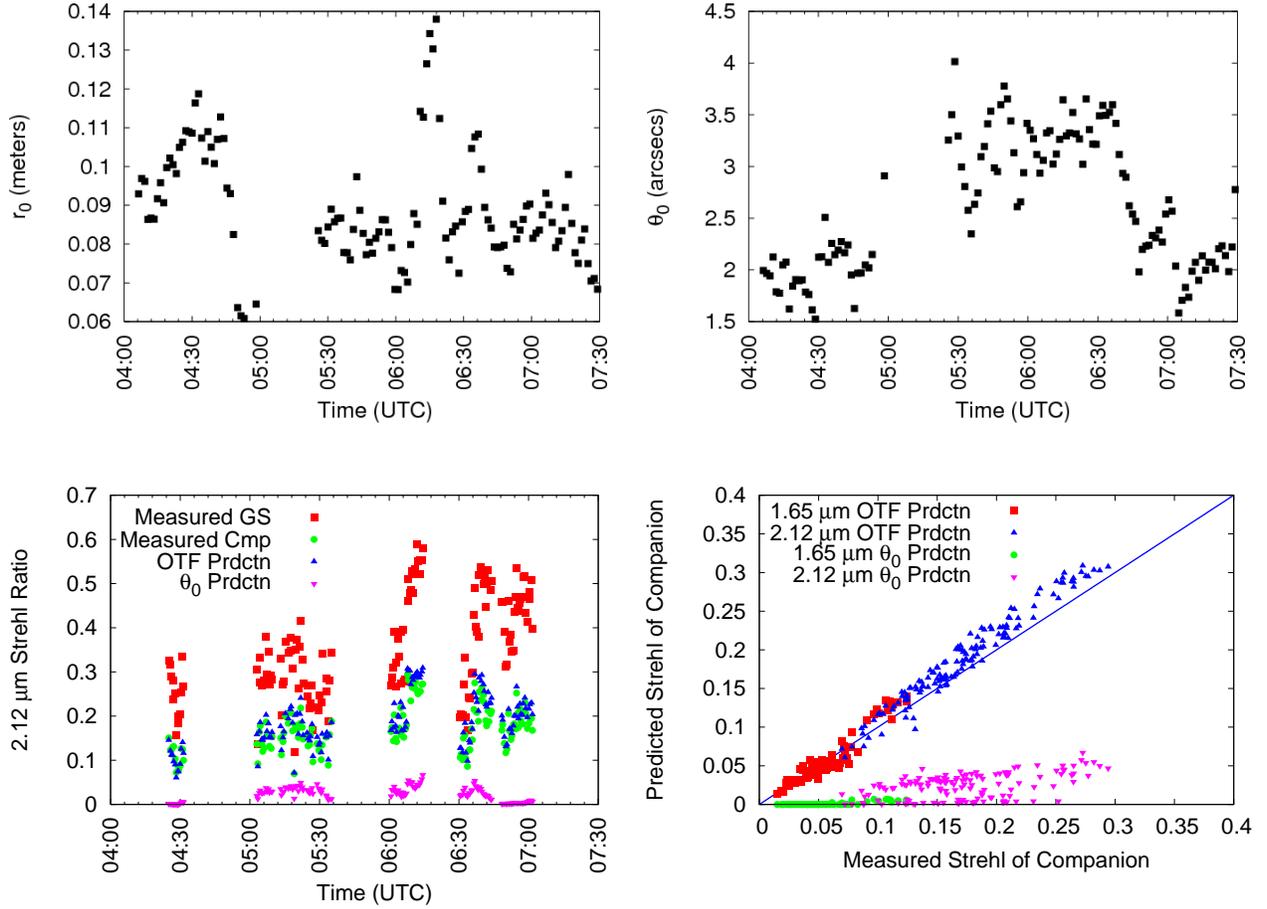}
\end{center}
\figcaption{Turbulence parameters and Strehl ratios over the course of
the observations.  The upper panels show the Fried parameter and
isoplanatic angle at .5 $\mu$m computed for the zenith angle of the
guide star, which ranged from 30 to 40 degrees over the three hour
observation.  The lower left panel shows the measured 2.12 $\mu$m
Strehl ratios of the guide star and companion.  Also plotted are the
Strehl ratios predicted from the OTF formulation in Section 2, and
predicted using the $\theta_{0}$ approximation described in the text.
The lower right panel shows the predicted vs. measured Strehl ratio of
the binary companion.  Strehl ratios at both 2.12 $\mu$m and 1.65
$\mu$m are included in this plot.  These plots indicate that the OTF
formulation accurately predicts the Strehl ratio degradation that
arises from ansioplanatism.  See the electronic edition of the Journal
for a color version of this figure.
\label{strehl_fig}
}
\end{figure}

Figure \ref{strehl_fig} shows the values of $r_{0}$ and $\theta_{0}$
computed from the measured turbulence profiles that were acquired over
the course of the three hour observation.  These parameters have been
computed for the zenith angle of the guide star, which varied between
30 and 40 degrees during the observations.  The values of these
turbulence parameters display considerable temporal variation.  Figure
\ref{strehl_fig} also shows the time dependence of the Strehl ratios
at 2.12 $\mu$m.  These Strehl ratios have again been averaged over
five sequential exposures.  The measured guide star Strehls vary
significantly, and show little correlation with the Fried parameter.
This suggests that sources of wavefront error other than residual
fitting error contributed significantly to the guide star error
budget.  The measured Strehl ratios of the binary companion are also
plotted, along with the Strehl ratios computed from the predicted
companion OTF.  As in Figure \ref{obs_fig}, these predictions are in
excellent agreement with the measurements, and are able to track the
companion Strehl ratio to a few percent despite a factor of two
variability in the measured Strehl ratio of both the guide star and
the companion.  Also plotted are the Strehl ratios of the binary
companion predicted using a traditional error budget approach.  These
predictions were generated from the product of the guide star Strehl
ratio and the anisoplanatic Strehl ratio computed using the
$\theta_{0}$ approximation, as described in Section
\ref{aniso_numerical}.  These predictions fall well below the
measurements, as expected from the discrepancies in the anisoplanatic
Strehl ratio shown in Figure \ref{aniso_fig}.  Lastly, Figure
\ref{strehl_fig} contains a plot of measured vs. predicted 2.12 $\mu$m
and 1.65 $\mu$m Strehl ratios for the binary companion.  This plot
again illustrates strong agreement between the measured Strehl ratios
and those derived using the predicted companion OTF.

%%%%%%%%%%%%%%%%
%%% Figure 6 %%%
%%%%%%%%%%%%%%%%
\begin{figure}[t!]
\begin{center}
\includegraphics[width=\textwidth]{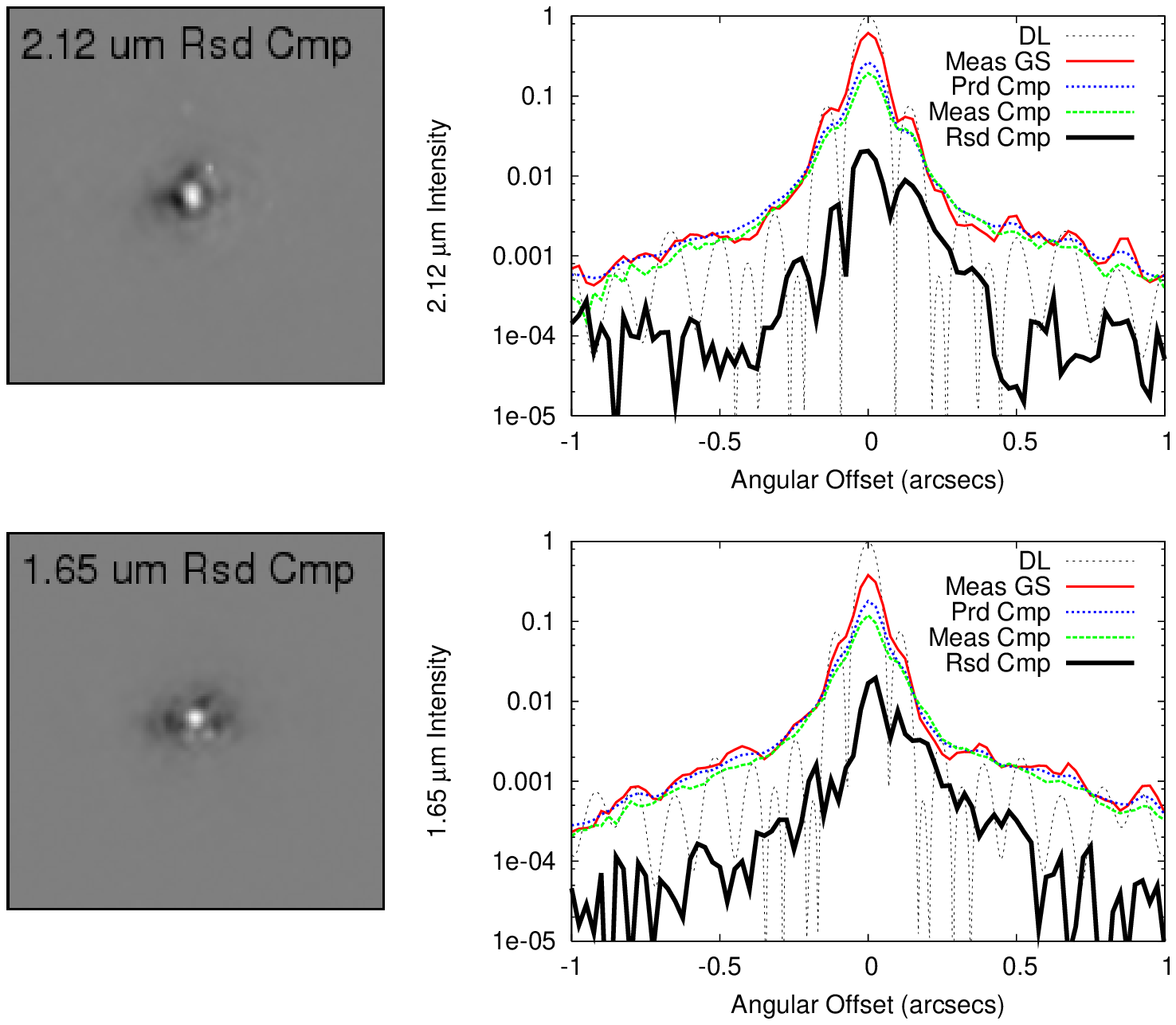}
\end{center}
\figcaption{Results of fitting the model of the companion PSF to the
observed data for the 2.12 $\mu$m and 1.65 $\mu$m observations shown
in Figure 4.  The left column shows the residuals in a 2 arcsec
field around the companion, displayed on a log stretch.  The right
column shows horizontal cuts through the simulated diffraction limited
PSF, the observed guide star PSF, and the predicted and observed
companion PSF.  The complex behavior displayed in the wings of the
diffraction limited PSF arises from light scattered by the struts
supporting the secondary mirror.  The magnitude of the residual
difference between the predicted and measured companion PSF is also
plotted.  At any point within the image, the predicted companion PSF
matches the observed data to an accuracy of about 10 percent.  See the
electronic edition of the Journal for a color version of this figure.
\label{resid_obs_fig}
}
\end{figure}

Results from fitting the predicted companion PSF to the data are shown
in Figure \ref{resid_obs_fig} for the observations shown in Figure
\ref{obs_fig}.  This figure shows that the prediction of the companion
PSF slightly overestimates the core and underestimates the wings of
the observed PSF.  Horizontal cuts through the simulated diffraction
limited PSF display complex morphology arising from light scattered by
the four struts supporting the secondary mirror.  The scattered light
is most pronounced along the horizontal and vertical axes of the
image, which are aligned with the support struts. Horizontal cuts
through the residuals indicate that the predicted companion PSF agrees
with the observations to about 10\% accuracy at any point in the
image.  Stated another way, this PSF fitting procedure has improved
the dynamic range of the observation by an order of magnitude.

These residuals display two regimes, in which the accuracy of the
predictions was limited by different effects.  At radii less than
about .5 arcsecs, the residuals were dominated by systematic errors
between the model and the observed data.  These systematics may arise
from a number of different effects.  The power spectrum of atmospheric
turbulence may be non-Komolgorov, so that Equation \ref{aniso_strfn}
is only approximately correct.  Accuracy in the turbulence profile
measurements from the DIMM/MASS unit may limit the quality of the
prediction.  Finally, nonlinear response or charge diffusion in the
infrared detector in PHARO may generate differential errors between
the measured guide star and companion PSFs.  At an angular separation
of about .5 arcsecs, the signal level in the residuals dropped below a
noise floor.  This noise floor is set by a combination of the detector
read noise, shot noise and quantization noise in the analog to digital
conversion.  The dominant source of noise depends on the efficacy of
the lowpass spatial filtering that was performed on the OTF, which
itself depends on the degree to which the images were oversampled.
Further investigation will be required to determine the noise source
that limits the precision of this technique in these two regimes.

At a radius of .5 arcsecs the value of the residual intensity was 3
to 4 orders of magnitude less than the peak of the measured companion.
This level of PSF rejection is comparable to that achieved in near
infrared AO observations that employ a Lyot coronograph.  These
systems provide about 4 to 5 orders of magnitude rejection at an
offset of 1'' from a star \citep{2000IAUS..202E..34O,
2005AN....326..952N}.  In comparing Lyot coronography to this fitting
technique, a disadvantage of the latter is that it requires the use of
another star to serve as the PSF reference.  On the other hand, Lyot
coronographs employ an opaque focal plane mask that limits the inner
working radius of the observation.  For example, the Lyot coronograph
in PHARO has masks with diameters of .41 and .91 arcsecs.  There is
no inner working radius in this fitting technique.

%%%%%%%%%%%%%%%%
%%% Figure 7 %%%
%%%%%%%%%%%%%%%%
\begin{figure}[t!]
\begin{center}
\includegraphics[width=\textwidth]{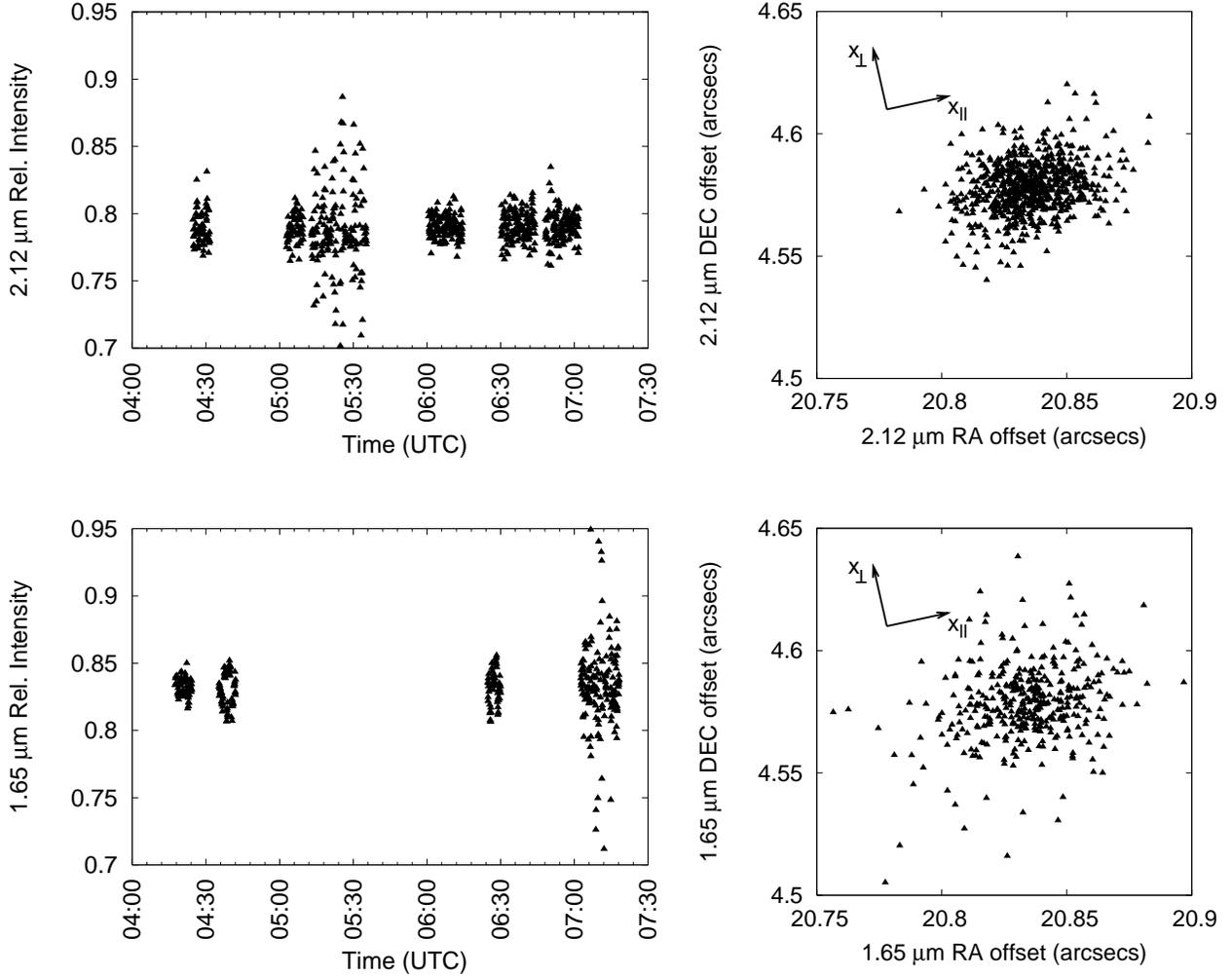}
\end{center}
\figcaption{Differential photometric and astrometric measurements of
the binary HD164984+HD164983 for 2.12 $\mu$m (upper) and 1.65 $\mu$m
(lower) observations.  Each point in these plots corresponds to a
single 2.8 second exposure at 2.12 $\mu$m, or 1.4 second exposure at
1.65 $\mu$m.  The left panels show the measured flux ratio of the
companion to primary throughout the course of the observations.  The
right panels show the measured differential angular offset between the
binary members for each exposure in the experiment.  The elliptical
scatter in the astrometric data arises from differential atmospheric
tilt jitter between the two stars that is induced by atmospheric
turbulence.  This tilt jitter is predicted to be larger along the
orientation of the binary, which lies along the $x_{\parallel}$ axis
indicated in these plots.
\label{phot_ast_fig}
}
\end{figure}

\begin{deluxetable}{rrrrrrrrr}
\tablewidth{0pt}
\tablecaption{Differential photometry and astrometry for the binary system HD164984+HD164983.
\label{phot_ast_table}
}
\tablehead{
\colhead{$\lambda$}                 & 
\colhead{Differential}              & 
\colhead{$\sigma_{\parallel}$}      & 
\colhead{$\sigma_{\perp}$}          &
\colhead{$\rho$}                    &
\colhead{P.A.}                      
\\
\colhead{($\mu$m)}                  & 
\colhead{Photometry}                & 
\colhead{(arcsec)}                  & 
\colhead{(arcsec)}                  &
\colhead{(arcsec)}                  &
\colhead{(deg)}                     
}
\startdata
2.12 & .7903$\pm$.0004 & .0152 & .0100 & 21.3322$\pm$.0006 & 282.3923$\pm$.0011 \\
1.65 & .8316$\pm$.0007 & .0228 & .0154 & 21.3306$\pm$.0012 & 282.3932$\pm$.0023 \\
\enddata
\end{deluxetable}

Figure \ref{phot_ast_fig} shows the differential photometry between
the binary members derived from fitting the predicted companion PSF to
the observed data.  The timeseries of differential photometric
measurements shows periods of substantial variability at 05:30 UT for
the 2.12 $\mu$m observations, and at 07:15 UT for the 1.65 $\mu$m
observations.  The origin of this variability is not clear, but could
plausibly be ascribed to cirrus.  When these data were excluded, the
standard deviation of the measured differential photometry was 1 part
in $10^{2}$ for both the 2.12 $\mu$m and 1.65 $\mu$m exposures.  The
mean differential photometry between the binary members is shown in
column 2 of Table 1.  The uncertainties quoted on these measurements
are the estimated errors of the mean \citep{1992drea.book.....B}, and
are less than 1 part in $10^{3}$.

This level of photometric stability may be compared to recent results.
\cite{2004SPIE.5490..504R} obtained differential photometric stability
of order 1 part in $10^{2}$ in J, H, and K band adaptive optics
observations of several binary systems on the Advanced Electro-Optical
System (AEOS) 3.6 m telescope.  Seeing limited observations in the
optical at the 6.5 m Multiple Mirror Telescope (MMT)
\citep{2005AJ....130.2241H} and in the near infrared at the 3.8 m
United Kingdom Infra-Red Telescope (UKIRT) \citep{2005MNRAS.363..211S}
displayed differential photometric stability of 1 part in $10^{3}$.
All of these observations employed broader filters and lower total
integration times than were used in this experiment.  Relative to
seeing limited observations, the sensitivity gains provided by
adaptive optics have afforded this level of photometric stability at
much lower flux levels, allowing observations of fainter targets.
This technique may be usefully applied to observations of eclipsing
binaries, transiting planets, and other systems that display
photometric variability in the near infrared.

Figure \ref{phot_ast_fig} also shows the differential astrometry
derived from the fit.  Each exposure yielded a single measurement, all
of which were combined to form the scatter plots in this figure.  The
astrometric offsets display an elliptical scatter, with larger errors
along the axis connecting the binary members.  This behavior is
consistent with that of differential atmospheric tilt jitter, which
arises from anisoplanatism of the tilt component of atmospheric
turbulence.  This effect leads to a random, achromatic fluctuation in
the relative displacement of two objects.  The standard deviation of
this tilt jitter differs along the axes parallel and perpendicular to
the orientation of the binary.  A three term approximation to the
parallel and perpendicular components of the variance arising from
differential atmospheric tilt jitter is given by
\cite{1994ewpt.book.....S}.
\begin{equation}
\label{tilt_jitter}
\left[
\begin{array}{c}
\sigma_{\parallel}^{2} \\
\sigma_{\perp}^{2}
\end{array} 
\right]
= 
2.67 {\mu_{2} \over D^{1/3}} \left({\theta \over D}\right)^{2} 
\left[
\begin{array}{c}
3 \\
1
\end{array} 
\right] 
-
3.68 {\mu_{4} \over D^{1/3}} \left({\theta \over D}\right)^{4} 
\left[
\begin{array}{c}
5 \\
1
\end{array} 
\right]
+
2.35 {\mu_{14/3} \over D^{1/3}} \left({\theta \over D}\right)^{14/3} 
\left[
\begin{array}{c}
17/3 \\
1
\end{array}
\right]
\end{equation}
Here the turbulence moments $\mu_{m}$ are defined as 
\begin{equation}
\mu_{m} = \int_{0}^{\infty} dz C_{n}^{2}(z) z^{m}
\end{equation}
Using the mean turbulence profile over the three hour observation, the
parallel and perpendicular components of differential atmospheric tilt jitter were
computed from Equation \ref{tilt_jitter} to be $\sigma_{\parallel} =
36$ mas and $\sigma_{\perp} = 21$ mas.  These values are somewhat
larger than those measured from the astrometric data, which appear in
columns 3 and 4 of Table 1.  The discrepancy likely arises from the
finite integration time of the exposures.  Assuming a characteristic
wind speed of 5 m/s, these integration times are comparable to the
wind crossing time for the 5 meter aperture, suggesting that tilt
jitter has partially averaged away.  This hypothesis is further
supported by the reduction in tilt jitter between the 1.4 second
exposures at 1.65 $\mu$m and the 2.8 second exposures at 2.12 $\mu$m.

The mean differential astrometry between the binary members is shown
in columns 5 and 6 of Table 1.  A PHARO pixel scale of .02522
asec/pixel was used in the calculation \citep{2004ApJ...617.1330M}.  The
uncertainties quoted for these measurements are again the estimated
errors of the mean.  This uncertainty is about 1 mas in the binary
separation $\rho$, and a few arcsecs in position angle.  Note that the
differential astrometry of the 2.12 $\mu$m and 1.65 $\mu$m
observations are in agreement at about the 1$\sigma$ level.  This
provides an independent validation of the accuracy of these
measurements.

This level of accuracy may be compared to recent astrometric results
obtained with the adaptive optics system on AEOS.
\cite{2004SPIE.5490..504R} performed I band adaptive optics
observations of a number of binaries and measured the differential
astrometry to accuracies of 10 to 20 mas.  These binaries had
separations up to 5 arcsecs.  \cite{2003ApJ...585.1007D} reported H
band observations of a 400 mas binary with an astrometric precision of
1 mas.  Note that to leading order the tilt variance in Equation
\ref{tilt_jitter} scales as $\theta^{2}/D^{7/3}$, so that these
binaries suffer much less from differential atmospheric tilt jitter
than the 21 arcsec binary observed in this experiment.  Compared to
the observations at Palomar reported here, both of these experiments
employed broader filters and lower total integration times.

These results may also be compared to those obtained using other
astrometric techniques.  An astrometric accuracy of order 100 $\mu$as
has been achieved on the Palomar Testbed Interferometer (PTI) for a 30
arcsec binary \citep{2000SPIE.4006..452L}, and recently accuracies of
tens of $\mu$as have been obtained for binaries with separations less
than an arcsec \citep{2004ApJ...601.1129L, 2006A&A...446..723M}.  This
instrument has a limiting magnitude of about 6 due to the 40 cm size
of its apertures.  The STEPS program on the Hale 5m telescope at
Palomar has achieved astrometric accuracies of less than 1 mas using
seeing limited observations in the visible.
\citep{1996ApJ...465..264P}.  These observations employed a 2
arcminute field of view and imaged crowded fields so as to establish
an astrometric reference grid from the multiple objects in the field.
Both of these techniques employ much broader filters than were used in
the observations reported here.  Further experiments will be required
to understand the circumstances under which the astrometric technique
described in Section \ref{data_analysis} is competitive with these
other methods.  Generally speaking, the sensitivity improvements
afforded by adaptive optics on a 5 meter aperture will permit
application of this methodology to fainter limiting magnitudes than
are accessible with PTI or STEPS, substantially increasing the number
of accessible targets.  Differential atmospheric tilt jitter scales as
$D^{-7/6}$, and application of this technique on a larger aperture
telescope will reduce this jitter while at the same time increasing
the signal to noise ratio and decreasing the width of the PSF core.
This would provide substantial improvements on the astrometric
accuracies that have been reported in this experiment.

\section{Conclusions}

The research presented here has drawn together a number of different
elements in order to generate predictions of the adaptive optics PSF.
Factorization of the OTF in Equation \ref{factored_otf} permits one to
use the guide star PSF as a reference for observations throughout the
field of view.  This PSF encapsulates the complex behavior of the
adaptive optics system that is otherwise very difficult to model.  The
covariance expression in Equation \ref{covariance} provides an
analytic formulation that captures the dependencies of anisoplanatism
on aperture diameter, observing wavelength, turbulence profile,
angular offset and zenith angle.  Measurements of the turbulence
profiles from the DIMM/MASS equipment provide the one input parameter
for these predictions that is not determined directly from the
observations.  These three elements provide a methodology for adaptive
optics PSF prediction that is accessible to direct experimental
validation.

The binary star observations described above are in excellent
agreement with these predictions.  The Strehl ratios computed from the
predicted companion PSFs match the measured values to an accuracy of a
few percent, despite a factor of two temporal variability in both the
guide star and companion Strehl ratios.  The predicted companion PSF
matches observations to about 10\% out to radii of 1 arcsec.  While
this agreement serves to validate the predictive methodology, it must
be emphasized that the adaptive optics PSF depends on a large number
of parameters.  This three hour experiment has tested these
predictions over a very modest region of the underlying parameter
space.  A broader application of this methodology at shorter observing
wavelengths, over wider fields of view, and under diverse turbulence
conditions will provide a more stringent test.  In some observing
conditions these predictions will almost certainly fail due to the
approximations discussed in Section \ref{theory}.  Likewise, the
target of this experiment was a relatively bright binary whose members
are of nearly identical magnitude.  Additional observations will be
required to understand the degree to which the photometric and
astrometric precision and the contrast levels reported here are
attainable in more diverse observational programs.  Experiments at
Palomar and Keck Observatories are currently being planned to perform
this experiment on more binaries, and to employ this methodolgy on
crowded field image data using both natural and laser guide star AO
systems.

Agreement between the predicted and observed companion PSF indicate a
level of consistency between turbulence profile measurements from the
DIMM/MASS equipment and the effects of anisoplanatism on the AO
compensated image quality.  Both Multiaperture Scintillation
Spectrometry and anisoplanatism are sensitive to higher altitude
turbulence, but each is sensitive to the underlying turbulence
statistics in a different way.  The former is sensitive to aberrations
introduced by short spatial wavelengths, which give rise to
scintillation.  In contrast, anisoplanatism arises from aberrations at
all spatial wavelengths, though the manner in which these aberrations
contribute depends on the altitude of the turbulence and the spatial
frequency of the aberration.  In this way, these two types of
measurements are sampling different spatial frequency ranges of the
turbulence power spectrum.  At some level the consistency of the
predictions and observations serve to validate the assumption of a
Komolgorov turbulence spectrum.  Further efforts will be required to
understand the level of agreement implied by these and future results.
One possible approach is to perform a sensitivity analysis by
generalizing Equation \ref{covariance} to include other classes of
power spectra.  An analysis of this type has been carried out by
\cite{2002A&A...382.1125L}, who considered the effects of
non-Komolgorov power spectra on differential atmospheric tilt jitter.

These results suggest several lines of longer term development that
may be of direct benefit to a number of astronomical applications.
The plot of Strehl ratio vs. guide star offset shown in Figure
\ref{aniso_fig} illustrates that the benefits of adaptive compensation
occur over fields much larger than anticipated from the $\theta_{0}$
approximation.  This suggests that adaptive optics systems with fields
of view of order several arcminutes may be usefully employed for near
infrared observations.  One of the challenges in interpreting these
data over wide fields arises from the temporal and field dependent
evolution of the adaptive optics PSF.  The methodology described above
allows a quantitative analysis of such wide field observations that
can account for these effects.  The photometric and astrometric
results presented in Section \ref{results} and the dynamic range
improvements implied by Figure \ref{resid_obs_fig} serve as
illustrations of the astronomical potential of this methodology.
Application of this technique to the imaging and deconvolution of
crowded fields and extended objects would constitute a natural
progression.  The approach could readily be applied to the
deconvolution of both image data and spatially resolved spectra
acquired with an integral field unit.

Lastly, agreement between predictions and observations serve as an
important on-sky validation that anisoplanatism is accurately
understood in the context of near infrared astronomical observations
over arcminute fields of view.  Anisoplanatism constitutes the
fundamental process underlying the use of tomography in adaptive
optics.  \cite{tyler94} describes the application of the phase
covariance in Equation \ref{covariance} in combining wavefront
measurements from multiple guide stars to form an estimate of the
wavefront in a different direction.  That this same expression has
been used in this research to accurately predict the anisoplanatic
degradation of the near infrared adaptive optics PSF over an arcminute
field is a strong indication that tomographic algorithms will be
successful in the same observational context.

\acknowledgments

The author would like to thank many colleagues who participated in
discussions of this research, including Brent Ellerbroek, Richard
Dekany, Mitchell Troy, Andrei Tokovinin, Keith Taylor, Andrew Pickles,
Roger Smith, Don Gavel and Matthias Schoeck.  The author gratefully
acknowledges the efforts of the Thirty Meter Telescope site testing
group and the Palomar Observatory staff in setting up and maintaining
the turbulence monitoring equipment at Palomar Observatory.

This paper was prepared as part of the work of the Thirty Meter
Telescope (TMT) Project. TMT is a partnership of the Association of
Universities for Research in Astronomy (AURA), the Association of
Canadian Universities for Research in Astronomy (ACURA), the
California Institute of Technology, and the University of
California. The partners gratefully acknowledge the support of the
Gordon and Betty Moore Foundation, the U.S. National Science
Foundation (NSF), the National Research Council of Canada, the Natural
Sciences and Engineering Research Council of Canada, and the Gemini
Partnership.

This work has also been supported by the National Science Foundation
Science and Technology Center for Adaptive Optics, managed by the
University of California at Santa Cruz under cooperative agreement
No. AST - 9876783.

{\it Facilities:} \facility{Hale}

\appendix

\section{Three Familiar Results in Adaptive Optics}

This Appendix uses Equation \ref{covariance} to recover three familiar
results in adaptive optics.  These results serve to illustrate the
validity of Equation \ref{covariance}, and its broad applicability in
performing analytic and numerical calculations in adaptive optics.
Equation \ref{covariance} represents the piston removed phase
covariance on a circular aperture in the presence of Komolgorov
turbulence, and the three results below are valid under these
assumptions.

\subsection{Phase Structure Function for Uncompensated Turbulence}
As a first example, consider the phase structure function
$D_{\phi}\left(\vec{r}_{1},\vec{r}_{2}\right)$ in the presence of
uncompensated turbulence.
\begin{eqnarray}
D_{\phi}\left(\vec{r}_{1},\vec{r}_{2}\right) & = & \left\langle\left[\phi\left(\vec{r}_{1}\right)-\phi\left(\vec{r}_{2}\right)\right]^{2}\right\rangle \\
& = & \left\langle\left[\phi\left(\vec{r}_{1}\right)\right]^{2}\right\rangle + \left\langle\left[\phi\left(\vec{r}_{2}\right)\right]^{2}\right\rangle - 
2  \left\langle\phi\left(\vec{r}_{1}\right)\phi\left(\vec{r}_{2}\right)\right\rangle \nonumber
\end{eqnarray}
The three covariance functions may be rewritten in terms of Equation
\ref{covariance}.  In the resulting expression, all dependencies on
the functions $G_{1}$ and $G_{2}$ drop out in the difference, leaving
only
\begin{eqnarray}
D_{\phi}\left(\vec{r}_{1},\vec{r}_{2}\right) & = & 
2^{8/3} \Xi k^{2} 
\left\vert\left(\vec{r}_{1} - \vec{r}_{2}\right)\right\vert^{5/3} \int dz\; C_{n}^{2}(z)
\end{eqnarray}
Define the Fried parameter $r_{0}$ as
\begin{equation}
r_{0}^{-5/3} = 2^{8/3} {\Xi \over \Lambda} k^{2} \int_{0}^{\infty}dz \; C_{n}^{2}(z)
\end{equation}
where the constant $\Lambda$ is
\begin{equation}
\Lambda = 2 \left[{24 \over 5}\;\Gamma\left({6 \over 5}\right)\right]^{5/6} = 6.88388
\end{equation}
The phase structure function may be rewritten in terms of $r_{0}$ as
\begin{equation}
D_{\phi}\left(\vec{r}_{1},\vec{r}_{2}\right)  = \Lambda\left({\left\vert\vec{r}_1 - \vec{r}_{2}\right\vert \over r_{0}}\right)^{5/3}
\end{equation}
This is the well known expression for the piston removed phase
structure function on a circular aperture in the presence of
Komolgorov turbulence.

\subsection{Aperture Averaged Phase Variance for Uncompensated Turbulence}
As a second example, the aperture averaged phase variance in the
presence of uncompensated turbulence is evaluated.  This quantity may
be computed by integrating the phase variance
$\left\langle\left[\phi(\vec{r})\right]^{2}\right\rangle$ over the
circular aperture and dividing by the area of the aperture.  The phase
variance may be rewritten using Equation \ref{covariance}, yielding

\begin{equation}
\label{apave_phasevariance_eqn}
{4 \over \pi D^{2}} \int d\vec{r} \; \left\langle\left[\phi(\vec{r})\right]^{2}\right\rangle = {\Lambda \over 2^{8/3}}\left({D \over r_{0}}\right)^{5/3}
\left\{\left[{8 \over \pi D^{2}} \int d\vec{r} G_{1}\left(\left\vert{2 \over D} \vec{r}\right\vert\right)\right] - G_{2}\left(0\right)\right\}
\end{equation}
The integral over the hypergeometric function in $G_{1}$ may be performed term by term, yielding
\begin{eqnarray}
{8 \over \pi D^{2}} \int d\vec{r} G_{1}\left(\left\vert{2 \over D} \vec{r}\right\vert\right) & =  &
{12 \over 11} \vphantom{F}_{2}F_{1}\left(-{11 \over 6}, -{5 \over 6}; 2; 1\right) \\
& = & {12 \over 11} {\Gamma(2)\Gamma({14 \over 3}) \over \Gamma({23 \over 6}) \Gamma({17 \over 6})} \nonumber
\end{eqnarray}
where the second equality has employed the relationship \citep{1972hmf..book.....A}
\begin{equation}
\vphantom{F}_{2}F_{1}\left(a, b; c; 1\right) = {\Gamma(c)\Gamma(c - a - b) \over \Gamma(c-a) \Gamma(c-b)}
\end{equation}
The second term in Equation \ref{apave_phasevariance_eqn} may be written
\begin{eqnarray}
G_{2}\left(0\right) & = &
{2^{11/3} \over \pi} \int_{0}^{1} dy y^{8/3} \left[\cos^{-1} y - y\sqrt{1 - y^{2}}\right] \\
& = & {9 \over 187}{2^{28/3} \over \pi} {\left[\Gamma({7 \over 3})\right]^{2} \over \Gamma({14 \over 3})} \nonumber
\end{eqnarray}
The integral has been evaluated analytically using the substitution $y
= \cos\psi$.  Further manipulations involving relationships between
Gamma functions or direct numerical calculation show that the first
term evaluates to exactly twice the second.  The resulting expression
becomes
\begin{eqnarray}
{4 \over \pi D^{2}} \int d\vec{r} \; \left\langle\left[\phi(\vec{r})\right]^{2}\right\rangle & = &
{6 \over 11} {\Gamma(2)\Gamma({14 \over 3}) \over \Gamma({23 \over 6}) \Gamma({17 \over 6})} {\Lambda \over 2^{8/3}} \left({D \over r_{0}}\right)^{5/3} \\
& = & 1.03242 \left({D \over r_{0}}\right)^{5/3} \nonumber
\end{eqnarray}
This is the well known result for the aperture averaged phase variance
in the presence of uncompensated turbulence.

\subsection{Aperture Averaged Residual Phase Variance from Anisoplanatism}
As a final example of particular relevance to this paper, consider the
aperture averaged residual phase variance due to anisoplanatism.  This
quantity may be computed by integrating the residual phase variance
$\left\langle\left[\phi_{b}(\vec{r}) - \phi_{a}(\vec{r})\right]^{2}\right\rangle$
over the circular aperture and dividing by the area of the aperture.
Again using Equation \ref{covariance}, we find
\begin{eqnarray}
\label{aniso_eqn}
{4 \over \pi D^{2}} \int d\vec{r} \; \left\langle\left[\phi_{b}(\vec{r}) - \phi_{a}(\vec{r})\right]^{2}\right\rangle & = &
{4 \over \pi D^{2}} \int d\vec{r} \; \left[\left\langle\left(\phi_{b}(\vec{r})\right)^{2}\right\rangle + \left\langle\left(\phi_{a}(\vec{r})\right)^{2}\right\rangle - 
\right. \\
& & 
\left.2 \left\langle\phi_{a}(\vec{r})\phi_{b}(\vec{r})\right\rangle\right] \nonumber \\
& = & 2 \Xi k^{2} D^{5/3} \int_{0}^{\infty} dz \; C_{n}^{2}(z) 
\left\{
\left\vert\vec{\Omega}_{ab}(z)\right\vert^{5/3} + 
\right. \nonumber \\
& & 
G_{2}\left(\left\vert\vec{\Omega}_{ab}(z)\right\vert\right) - G_{2}(0)  + \nonumber \\
& & 
{4 \over \pi D^{2}} \int d\vec{r} 
\left[
G_{1}\left(\left\vert{2 \over D} \vec{r}\right\vert\right) - G_{1}\left(\left\vert{2 \over D} \vec{r} + \vec{\Omega}_{ab}(z)\right\vert\right) + 
\right.
\nonumber \\
& &
\left.\left.
G_{1}\left(\left\vert{2 \over D} \vec{r}\right\vert\right) - G_{1}\left(\left\vert{2 \over D} \vec{r} - \vec{\Omega}_{ab}(z)\right\vert\right)
\right]\right\}
\nonumber 
\end{eqnarray}

The first term is
\begin{eqnarray}
2 \Xi k^{2} D^{5/3} \int_{0}^{\infty} dz \; C_{n}^{2}(z) 
\left\vert\vec{\Omega}_{ab}(z)\right\vert^{5/3} 
& = & 
2^{8/3} \Xi k^{2} \left\vert\vec{\theta}_{ab}\right\vert^{5/3}\int_{0}^{\infty} dz \; C_{n}^{2}(z) z^{5/3} \\
& = & \left(\left\vert\vec{\theta}_{ab}\right\vert \over \theta_{0}\right)^{5/3} \nonumber
\end{eqnarray}
where the isoplanatic angle $\theta_{0}$ is defined as
\begin{equation}
\theta_{0}^{-5/3} = 2^{8/3} \Xi k^{2} \int_{0}^{\infty} dz \; C_{n}^{2}(z) z^{5/3}
\end{equation}
This is the well known isoplanatic angle approximation to the aperture
averaged residual phase variance due to anisoplanatism.

Consider expanding the remaining three pairs of terms in Equation
\ref{aniso_eqn} in a Taylor series about $\vec{\Omega}_{ab}(z) = 0$.
The constant terms cancel in the differences.  The linear term in
$\vec{\Omega}_{ab}(z)$ vanishes in the function
$G_{2}\left(\left\vert\vec{\Omega}_{ab}(z)\right\vert\right)$, since
it is a function only of the magnitude of this vector.  The linear
term in $\vec{\Omega}_{ab}(z)$ also vanishes in the sum of the terms
$G_{1}\left(\left\vert 2 \vec{r} / D -
\vec{\Omega}_{ab}(z)\right\vert\right)$ and $G_{1}\left(\left\vert 2
\vec{r} / D + \vec{\Omega}_{ab}(z)\right\vert\right)$, since this
vector enters these two terms with opposite sign.  Thus, the remaining
terms in Equation \ref{aniso_eqn} have a leading order dependence of
$\left\vert\vec{\Omega}_{ab}(z)\right\vert^{2}$.  These terms are only
slightly higher order than the first term in this equation, and their
contribution to the aperture averaged residual phase variance can be
significant even at modest angular offsets.  This is illustrated by
the large discrepancy between the isoplanatic angle approximation to
the aperture averaged residual phase variance and the exact result
shown in Figure \ref{aniso_fig}.


\begin{thebibliography}{}
\bibitem[Abramowitz \& Stegun(1972)]{1972hmf..book.....A} Abramowitz, M., 
\& Stegun, I.~A., Handbook of Mathematical Functions, New York: 
Dover, 1972
\bibitem[Beckers(1988)]{1988vltt.conf..693B} Beckers, J.~M.\ 1988, Very 
Large Telescopes and their Instrumentation, ESO Conference and Workshop 
Proceedings, Proceedings of a ESO Conference on Very Large Telescopes and 
their Instrumentation, Garching: 
European Southern Observatory (ESO), 1988, ed. Marie-Helene Ulrich., 
p.693, 693 
\bibitem[Bevington \& Robinson(1992)]{1992drea.book.....B} Bevington,
P.~R., \& Robinson, D.~K.\, Data Reduction and Error Analysis for the
Physical Sciences, New York: McGraw-Hill, 1992
\bibitem[Bracewell(1986)]{1986ftia.book.....B} Bracewell, R.~N., The
Fourier Transform and Its Applications, New York: McGraw-Hill, 1986
\bibitem[Christou et al.(2004)]{christou04} Christou, J.~C., 
Pugliese, G., K{\"o}hler, R., \& Drummond, J.~D.\ 2004, \pasp, 116, 734 
\bibitem[Diolaiti et al.(2000)]{diolaiti00b} Diolaiti, E., 
Bendinelli, O., Bonaccini, D., Close, L., Currie, D., \& Parmeggiani, G.\ 
2000, \aaps, 147, 335 
\bibitem[Drummond et al.(2003)]{2003ApJ...585.1007D} Drummond, J., Milster, 
S., Ryan, P., \& Roberts, L.~C.\ 2003, \apj, 585, 1007 
\bibitem[Ellerbroek et al.(2005)]{2005SPIE.5903...20E} Ellerbroek, B., 
Britton, M., Dekany, R., Gavel, D., Herriot, G., Macintosh, B., \& Stoesz, 
J.\ 2005, \procspie, 5903, 20 
\bibitem[Flicker \& Rigaut(2005)]{flicker05} Flicker, R. \& Rigaut, F. 2005, J. Opt. Soc. Am. A 22, 504
\bibitem[Fusco et al.(2000)]{2000A&AS..142..149F} Fusco, T., Conan, J.-M., 
Mugnier, L.~M., Michau, V., \& Rousset, G.\ 2000, \aaps, 142, 149 
\bibitem[Goodman(1985)]{1985stop.book.....G} Goodman, J.~W.,
Statistical Optics, New York: Wiley, 1985
\bibitem[Hammer et al.(2004)]{2004SPIE.5382..727H} Hammer, F., et al.\ 
2004, \procspie, 5382, 727 
\bibitem[Hartman et al.(2005)]{2005AJ....130.2241H} Hartman, J.~D., Stanek, 
K.~Z., Gaudi, B.~S., Holman, M.~J., \& McLeod, B.~A.\ 2005, \aj, 130, 2241 
 \bibitem[Hayward et al.(2001)]{2001PASP..113..105H} Hayward, T.~L., Brandl, 
B., Pirger, B., Blacken, C., Gull, G.~E., Schoenwald, J., \& Houck, J.~R.\ 
2001, \pasp, 113, 105.
\bibitem[Kornilov et al.(2003)]{2003SPIE.4839..837K} Kornilov, V., 
Tokovinin, A.~A., Vozyakova, O., Zaitsev, A., Shatsky, N., Potanin, S.~F., 
\& Sarazin, M.~S.\ 2003, \procspie, 4839, 837 
\bibitem[Lane et al.(2000)]{2000SPIE.4006..452L} Lane, B.~F., Colavita, 
M.~M., Boden, A.~F., \& Lawson, P.~R.\ 2000, \procspie, 4006, 452 
\bibitem[Lane \& Muterspaugh(2004)]{2004ApJ...601.1129L} Lane, B.~F., \& 
Muterspaugh, M.~W.\ 2004, \apj, 601, 1129 
\bibitem[Lazorenko(2002)]{2002A&A...382.1125L} Lazorenko, P.~F.\ 2002, 
\aap, 382, 1125 
\bibitem[L{\'e}na \& Lai(1999)]{1999aoa..book..351L} L{\'e}na, P., \& Lai, 
O.\ 1999, Adaptive Optics in Astronomy, 351 
\bibitem[Metchev \& Hillenbrand(2004)]{2004ApJ...617.1330M} Metchev, S.~A., 
\& Hillenbrand, L.~A.\ 2004, \apj, 617, 1330 
 \bibitem[Muterspaugh et al.(2006)]{2006A&A...446..723M} Muterspaugh, M.~W., 
Lane, B.~F., Konacki, M., Burke, B.~F., Colavita, M.~M., Kulkarni, S.~R., 
\& Shao, M.\ 2006, \aap, 446, 723 
\bibitem[Nakajima et al.(2005)]{2005AN....326..952N} Nakajima, T., et al.\ 
2005, Astronomische Nachrichten, 326, 952 
\bibitem[Oppenheimer et al.(2000)]{2000IAUS..202E..34O} Oppenheimer, B.~R., 
Dekany, R., Hayward, T., Brandl, B., \& Troy, M.\ 2000, IAU Symposium, 202, 
\bibitem[de Pater et al.(2004)]{2004Icar..169..250D} de Pater, I., Marchis, 
F., Macintosh, B.~A., Roe, H.~G., Le Mignant, D., Graham, J.~R., \& Davies, 
A.~G.\ 2004, Icarus, 169, 250 
\bibitem[Pravdo \& Shaklan(1996)]{1996ApJ...465..264P} Pravdo, S.~H., \& 
Shaklan, S.~B.\ 1996, \apj, 465, 264
\bibitem[Roberts et al.(2004)]{2004SPIE.5490..504R} Roberts, L.~C., et al.\ 
2004, \procspie, 5490, 504 
\bibitem[Sasiela(1994)]{1994ewpt.book.....S} Sasiela, R.~J.,
Electromagnetic Wave Propagation in Turbulence, Berlin: Springer,
1994
\bibitem[Skidmore et al.(2004)]{2004SPIE.5489..154S} Skidmore, W., et al.\ 
2004, \procspie, 5489, 154 
\bibitem[Snellen(2005)]{2005MNRAS.363..211S} Snellen, I.~A.~G.\ 2005, 
\mnras, 363, 211 
\bibitem[Steinbring et al.(2002)]{steinbring02} Steinbring, E., et al.\ 2002, \pasp, 114, 1267 
\bibitem[Tokovinin et al.(2003)]{2003MNRAS.340...52T} Tokovinin, A., 
Baumont, S., \& Vasquez, J.\ 2003, \mnras, 340, 52 
\bibitem[Tokovinin et al.(2005)]{2005PASP..117..395T} Tokovinin, A., 
Vernin, J., Ziad, A., \& Chun, M.\ 2005, \pasp, 117, 395
\bibitem[Troy et al.(2000)]{2000SPIE.4007...31T} Troy, M., et al.\ 2000, 
\procspie, 4007, 31
\bibitem[Tyler(1983)]{tyler83} Tyler, G. 1983, J. Opt. Soc. Am. A 1, 251
\bibitem[Tyler(1994)]{tyler94} Tyler, G. 1994, J. Opt. Soc. Am. A 11, 409
\bibitem[Veran et al.(1997)]{veran97} Veran, J., Rigaut, F., Matre, H. \& Rouan, D. 1997, J. Opt. Soc. Am. A 14, 3057
\bibitem[Vernin \& Munoz-Tunon(1995)]{1995PASP..107..265V} Vernin, J., \& 
Munoz-Tunon, C.\ 1995, \pasp, 107, 265 
\bibitem[Voitsekhovich et al.(1998)]{1998A&AS..133..427V} Voitsekhovich, 
V.~V., Orlov, V.~G., Cuevas, S., \& Avila, R.\ 1998, \aaps, 133, 427 
\bibitem[Wei{\ss} et al.(2002)]{weiss2002} Wei{\ss}, A., Hippler, 
S., Kasper, M., Wooder, N., \& Quartel, J.\ 2002, ASP Conf.~Ser.~266: 
Astronomical Site Evaluation in the Visible and Radio Range, 266, 86 
 \end{thebibliography}
\end{document}